\documentstyle[prl,aps,amsmath,epsfig,floats,twocolumn]{revtex}

\begin{document}
\draft
\twocolumn[\hsize\textwidth\columnwidth\hsize\csname@twocolumnfalse%
\endcsname

\preprint{}

\title{Neutron Scattering Signature of
$\mathbf d$-density Wave Order in the Cuprates}
\author{Sudip Chakravarty, Hae-Young Kee, and Chetan Nayak}
\address{Department of Physics, University of California Los Angeles,  Los Angeles, California 
  90095--1547}
\date{\today}
\maketitle

\begin{abstract}
An ordered $d$-density wave (DDW) state has been
proposed as an explanation of the pseudogap
phase in underdoped high-temperature superconductors.
The staggered currents associated
with this order have signatures which
are qualitatively different from those of ordered spins.
We apply the order parameter theory to an
orthorhombic bilayer system and show that the expected
magnitude as well as the momentum, energy, and polarization
dependence of the consequent neutron 
scattering is consistent with the
findings of a recent experiment.
\end{abstract}

]
\narrowtext

\paragraph{Introduction.} The pseudogap phase of the
underdoped high-temperature superconducting
cuprates is notable in its departure from the
behavior of a conventional metal. Recently,
it was proposed\cite{CLMN,CK} that this unusual behavior
is due to a new broken symmetry, termed $d$-density wave (DDW)\cite{Nayak00}, which
competes with superconductivity. 
The crucial feature of this order is a staggered pattern of orbital
currents that spontaneously appear at the onset of the pseudogap
phase. The underdoped superconducting
state {\em coexists} with this order. 
This proposal 
of the existence of an {\em order parameter} is
fundamentally different from the proposal of staggered current
{\em fluctuations}\cite{Patrick}.

We discuss here the experimental signatures of this
unusual state vis-\`a-vis a recent neutron scattering
experiment in underdoped
YBCO$_{6.6}$\cite{Mook}.  This experiment has identified
rods of Bragg scattering (within the energy
resolution window of 1 meV) arising from small
magnetic moments  which increase in strength below the
pseudogap temperature, with a
further increase below the superconducting transition temperature.
The in-plane wavevector is ${\bf Q}=(\pi/a,\pi/a)$,
where $a$ is the in-plane lattice spacing,
and the intensity is at the level of the
background for perpendicular wavevector transfer $q_z=0$, implying an
``antiferromagnetic" coupling between the layers within a bilayer. Both of
these features are precisely the same as those observed in the undoped antiferromagnet, where the
signal is from ordered Cu moments. Such
scattering from ordered Cu moments deep in the
superconducting state would be surprising to say the least.

However, there are four important ways in
which these experimental results
differ from the corresponding ones for
the undoped antiferromagnet:
(1) the magnitude of the observed moments
is of order $2\times 10^{-2} \mu_B$
at 10 K, which is $50$ times smaller than
that observed in the undoped antiferromagnet;
(2) the intensity decreases rapidly with
scattering wavevector ${\bf q}$ in a manner
which is inconsistent with ordered
Cu spins;
(3) aside from the elastic
peak, there is no significant intensity up to 20 meV. In other words, there
is no evidence for the Goldstone modes (magnons) which
must be present if spin-rotational
symmetry is broken. (4) Finally, it is
found that there is no 3D order down to the lowest
temperatures studied despite the fact that the in-plane
spatial correlations are resolution limited, with
a correlation length greater than 200\AA.

We argue that these four are the most robust features
of the experiment and that they fit the explanation of
DDW order but are inconsistent with spin order. In order to understand the experiment, we first 
introduce the theoretical framework of the DDW. Next, we address the ratio of the
intensities for various Bragg reflections and then examine the ratio of the spin-flip to non-spin flip
scattering for polarized neutrons.

\paragraph{Currents associated with DDW order.}
The DDW order parameter\cite{Nayak00}
is a spin singlet particle-hole condensate given 
by the equal time correlation function ($\alpha$, $\beta$ are spin indices):
$
\left\langle {c^{\dagger}}_{{\bf k+Q}\alpha}
c_{{\bf k}\beta} \right\rangle =
i{\Phi_{\bf Q}}\,Y({\bf k})\,\,
{\delta_{\alpha\beta}},
$
where $\Phi_{\bf Q}$ is a real order parameter, and
$c_{\beta}$ is an electron destruction operator. The internal
degree of freedom of the condensate is given
by the
$d_{x^2-y^2}$ function
$
Y({\bf k})=\frac{1}{2}\left( \cos {k_x}a - \cos {k_y}a\right).
$
Charge density is not modulated in the DDW
because $\sum_{\bf k}Y({\bf k})=0$! It is actually the current
density that is modulated, as we discuss below.
A conventional charge density wave
would occur if the internal orbital degree of freedom
of the particle-hole condensate were of $s$-wave character, e.g.\ if 
$Y({\bf k})$ were a constant. The terminology is a
very convenient way of classifying the internal
orbital degree of freedom of a particle-hole condensate.  

Neutron scattering from the DDW state is
determined by its associated
current distribution. This distribution
depends on the details of the current paths
along which electrons move in a copper-oxide
superconductor.
As we discuss below, unpolarized neutron
scattering is fairly insensitive
to these details, but the ratio of non-spin-flip
to spin-flip polarized neutron scattering is
not.

The charge and spin distribution of an electron
at a Cu site is given by the well-known Cu
form factor (see below). However,  due to mixing of different orbitals, the profile
of current flow is more complicated; it will be sensitive to many-body
effects when the current is substantial, and it will
be affected by the bi-layer coupling and, in YBCO,
by the influence of the chains. We make a simple and tractable
{\em Ansatz} for this current distribution, including 
the effects of orthorhombicity
due to the chains. The
orthorhombicity, even though it is small in terms of the lattice
constants, is well known to give rise to remarkable 
anisotropy; for example the
$a-b$-anisotropy of the
superfluid density can be as large as
2-3\cite{Basov}, and perhaps even larger\cite{Dynes}.

Thus, we define the Fourier transform of the expectation value of the current,
$\langle {\bf j}({\bf q})\rangle$, 
satisfying ${\bf q} \cdot \langle {\bf j}({\bf q})\rangle = 0$, by
\begin{eqnarray}
\langle{\bf j}({\bf q})\rangle &\propto& \Phi_{\bf Q}
\sum_{{\bf G}_{\parallel}}\delta_{{\bf q}_{\parallel},{\bf G}_{\parallel}}\,
\,f({\bf q})\nonumber\\& & {\hskip -0.5 cm} \times 
\left[\left(\alpha({\bf q})\frac{\hat{\bf x}}{q_x} -
\beta({\bf q})\frac{\hat{\bf y}}{q_y}\right) -
(\alpha({\bf q})-\beta({\bf q}))\frac{\hat{\bf z}}{q_z}\right]
\label{eqn:j_q}
\end{eqnarray}
where ${\bf q}_{\parallel}=(q_x,q_y)$, ${\bf G}_{\parallel} = (G_x,G_y)$, and ${
\bf
G}=(2\pi H/a, 2\pi K/a, 2\pi L/c)$ is a magnetic reciprocal lattice vector, where the lattice
constants are
$a=3.86 \mbox{\AA}$ and $c=11.82 \mbox{\AA}$; we ignore slight
orthorhombicity in the lattice constants, but not in the 
physical quantities in which it is magnified.
Note that $f({\bf q})=\sin(\frac{q_z d}{2} )$,
where $d=3.25 \mbox{\AA}$ is the intrabilayer separation. The temperature and
doping dependence of the currents are determined by the DDW order parameter $\Phi_{\bf Q}$,
which must appear as a thermodynamic phase transition.

If the electrons behaved classically and took
straight line trajectories between point-like Cu
sites, we would have $\alpha({\bf q})=\beta({\bf q})=1$.
In reality, however, these are non-trivial functions
of ${\bf q}$. Guided by the phenomenology, we choose the simplest model in which these are non-trivial functions
of $q_z$ alone. At long scales, the current distribution
is essentially classical, $\alpha(0)=\beta(0)=1$.
At short distances, the thickness of the current
trajectories becomes apparent. We assume that
currents flowing along the $\hat{\bf y}$ axis
are more spread out, reflecting
the underlying orthorhombicity of YBCO,
$\alpha(q_z)\neq \beta(q_z)$
for ${q_z}$ large. In tetragonal materials,
the asymmetry could be  due to spontaneous symmetry
breaking\cite{Kivelson}. 

The requirement  $\alpha(0)=\beta(0)=1$ automatically
leads to zero average 
current along $\hat{\bf z}$ in the bilayer model
because of the bilayer form
factor $f({\bf q})$. If we want to interpret
$\alpha(q_z)$ and $\beta(q_z)$ to 
be merely form factors and want to adhere to the original 
order parameter\cite{CLMN,Nayak00}, we can do so
if we stipulate that $\alpha(q_z)-\beta(q_z)\sim q_z^2$, 
as $ q_z\to 0$.
This will make the average $z$-component of the
current vanish (as it must) for the single layer
model for which the same expression
for the current holds, except that the factor $f({\bf q})$ 
is missing. 

The orbital currents give rise to magnetic fields of order $10\, G$,
\cite{CLMN,Nayak00,Hsu} and a very small
magnetic moment,  of order $4\times 10^{-2} \mu_B$,
consistent with the experimental estimate of
order $2\times 10^{-2} \mu_B$ \cite{Mook}.

\paragraph{Unpolarized elastic neutron scattering.} 
We first recall spin-scattering and then discuss scattering from 
orbital currents. 
The  well-known  scattering cross section
of unpolarized neutrons
from localized Cu spins in YBCO, 
when the net spin moment is along the $c$-axis, is
\begin{equation}
\left( \frac{d\sigma}{d\Omega}
\right)_{\rm s}
\propto \sum_{{\bf G}_{\parallel}} \delta_{{\bf q}_{\parallel},{\bf G}_{\parallel}}\  g^2({\bf q})  
f^2({\bf q})\left(1-\frac{q_z^2}{q^2}\right),
\label{eq:perpspin}
\end{equation}
where $\bf q$ is the momentum transfer, and the atomic
form factor $g({\bf q})$ is the
Fourier transform of the normalized density of unpaired
electrons in a single ion\cite{Shamoto}.
We have assumed that the spins are
antiferromagnetically coupled within an YBCO bilayer
and that different bi-layers are uncoupled, so
that the scattering cross-section is proportional to that
due to a single bi-layer.

Similarly, there is a formula which applies
when the spin moment lies
in the $x-y$ plane\cite{Shamoto}. If the spin direction in
the plane is averaged over magnetic domains or
averaged due to spin glass order, this is 
\begin{equation}
\left( \frac{d\sigma}{d\Omega}
\right)_{\rm s}
\propto \sum_{{\bf G}_{\parallel}} \delta_{{\bf q}_{\parallel},{\bf G}_{\parallel}}\ g^2({\bf q})  
f^2({\bf q})\left(1 + \frac{q_z^2}{q^2}\right).
\label{eq:parallels}
\end{equation}
It is worth emphasizing that the quantity in large parentheses in
(\ref{eq:parallels}) increases with $q_z$, as does
its counterpart in the case in which the spins point in some direction
in the $a-b$ plane which is not averaged over.

Next, we consider Bragg scattering from orbital currents\cite{Hsu}.
The neutron magnetic moment generates a vector potential,
$\bf A$, given by
$
{\bf A} = {\boldsymbol\mu} \times
\frac{({\bf r}_e-{\bf r}_n)}{|{\bf r}_e-{\bf r}_n|^3},
$
where
${\boldsymbol\mu}= -1.91(\frac{e \hbar}{m_n }) {\bf s}_n$, ${\bf s}_n$ is
the neutron spin, and $m_n$ is the mass of a neutron;
the electron and the neutron
coordinates are 
${\bf r}_e$ and ${\bf r}_n$, respectively.
The coupling of the electrons to this gauge field
is given in momentum space by the Hamiltonian
$
H_{\rm int} = \int \frac{{d^3}{\bf q}}{(2\pi)^3}\:
{\bf j}({\bf q})\cdot {\bf A}({\bf q}).
$
The Bragg intensity of {\em unpolarized} neutrons is then 
\begin{eqnarray}
\left( \frac{d\sigma}{d\Omega}\right)_{\rm o}
&\propto&
\frac{\left|\langle{\bf j}({\bf q})\rangle\right|^2}{q^2}
\propto\sum_{{\bf G}_{\parallel}}f^2({\bf q}) \beta^2(q_z)
\frac{\delta_{{\bf q}_{\parallel},{\bf G}_{\parallel}}}{q^2}\nonumber \\ 
{\hskip -1.5 cm}&\times&
\left[\frac{\lambda^2(q_z)}{q_x^2} +
\frac{1}{q_y^2} +
\frac{\left(\lambda(q_z)-1\right)^2}{q_z^2}\right],
\label{eq:bragg_qz}
\end{eqnarray}
where $\lambda(q_z)=\alpha(q_z)/\beta(q_z)$.
We emphasize that there are no approximations in this formula, such as the SU(n)
 mean field
approximation \cite{Hsu}; it simply follows from the
assumption of DDW order.

We estimate the factor $\beta^2(q_z)$ by $g^2({\bf q})$ for Cu
spins found by Shamoto {\em et
al.}\cite{Shamoto} as an {\em upper bound}, because  $\beta^2(q_z)$ must fall off faster 
than  $g^2({\bf q})$ as the  charge distribution for orbital 
current loops is more spread out than the atomic orbitals. 
The case for orbital currents made below will be
even stronger with the true form factor.

The most salient feature distinguishing
(\ref{eq:perpspin}) and (\ref{eq:parallels}) from (\ref{eq:bragg_qz})
is apparent from dimensional analysis. The former
scale with ${\bf q}$ as $q^0$ while the latter dominantly
scale as $1/q^4$ (recall that $|{\bf j}({\bf q})|\sim 1/q$
according to (\ref{eqn:j_q}) and $\alpha(q_z)$ and $\beta (q_z)$
are of order unity).
The latter form follows from the
inherent spatial extent of a current loop and leads to a rapid
decay of the scattering cross-section with ${\bf q}$.

Let us consider the ratios of the Bragg scattering
intensities at various magnetic reciprocal lattice vectors;
we will quote the intensities at the representative
values, $\lambda(q_z)=1$ and $3$, thereby crudely replacing $\lambda(q_z)$
by a constant for wavevectors larger than or of order $2\pi/c$. From
(\ref{eq:bragg_qz}), we see that the ratio of the
intensity at $(0.5, 0.5, 1)$ (in reciprocal lattice units)
to that at $(1.5, 1.5, 1)$ is 111 at $\lambda=1$,
and $23$ at $\lambda=3$. Meanwhile
in the spin case, the same ratio is 1.9
if the spins lie in the $a-b$ plane 
and 1.4 if they are along the $c$-axis.
The strong suppression of the intensity at
$(1.5, 1.5, 1)$ in the orbital
case is consistent
with experiments\cite{Mook,Mookpc}
where no peak is observed at $(1.5, 1.5, 1)$ or
$(1.5, 1.5, 2)$, while peaks are
observed at $(0.5, 0.5, 1)$ and $(0.5, 0.5, 2)$.
The absence of an observable peak at $(1.5, 1.5, 1)$
is inconsistent with spin order.

Let us now contrast the ratios of the intensities
at different values of $q_z$ for spins pointing
within the plane, Eq.~\ref{eq:parallels}, and  orbital currents,
Eq.~\ref{eq:bragg_qz}, with the form factor correction
described above.
The ratio of the intensity at the
Bragg reflection $(0.5, 0.5, 1)$ to that at
$(0.5, 0.5, 2)$ is $0.96$ for orbital moments at $\lambda=1$,
increasing to $1.5$ at $\lambda=3$. For
spins pointing in the plane it is $0.5$.
Experimentally\cite{Mook} it is found to be
$1.16\pm0.16$, suggesting orbital currents as the source of scattering.
One can easily see from the known Cu form
factor\cite{Shamoto} that the correction due to
the atomic form factor for nearby Bragg reflections, such as
$(0.5,0.5,1)$ and $(0.5,0.5,2)$, is negligible. 

Moreover, for spin-moments lying
parallel to the plane, the ratio of the intensity
at $(0.5,0.5,5)$ to $(0.5,0.5,1)$ should be 1.6. This
is clearly not the case in the experiment\cite{Mook,Mookpc}
because there is no intensity above the background at
$(0.5,0.5,5)$. For orbital currents, the same ratio is 0.2
(which is only an upper bound) at $\lambda=1$ and 0.1
at $\lambda=3$; as a result, there may not be
a measurable intensity at $(0.5,0.5,5)$ as discussed in Ref.~\cite{Mook}, thus
confirming once again the source of the
scattering to be orbital currents.

\paragraph{Polarized elastic scattering.}
For polarized neutrons, we have
\begin{eqnarray}
\left( \frac{d\sigma}{d\Omega}\right)_{\alpha\to \beta}
&\propto& \frac{1}{q^4} \left|\langle\alpha|{\boldsymbol \mu}|\beta\rangle
\cdot{\bf q}\times\langle{\bf j}({\bf q})\rangle\,
\right|^2,
\end{eqnarray}
where $|\alpha\rangle$ and $|\beta\rangle$ are the initial
and final spin states
of the neutrons. It is easy to see that if the
neutron polarization is
parallel to the scattering vector, the 
scattering is entirely spin flip irrespective of
$\langle{\bf j}({\bf q})\rangle$.

In contrast, if it
is perpendicular to the scattering vector,
then the scattering can be either spin flip or
non-spin flip depending on $\langle{\bf j}({\bf q})\rangle$
and ${\bf q}$.
If  ${\bf q}=(H,H,L)$ and 
the neutrons are polarized perpendicular to ${\bf q}$,
in the $[1,{\overline 1},0]$ direction,
the ratio of the non-spin-flip to spin flip
scattering intensities is:
\begin{eqnarray}
\label{eqn:NSF/SF}
\frac{\text{NSF}}{\text{SF}} = 2\,
{\left(\frac{\lambda (q_z)-1}{\lambda(q_z)+1}\right)^2}
\:\:{\left(\frac{Hc}{La}\right)^2}
\left[1+\frac{1}{2}{\left(\frac{La}{Hc}\right)^2}\right]
\end{eqnarray}
While this ratio {\em vanishes} at $\lambda=1$,
it increases rapidly with $\lambda$ because
$c\gg a$. At
(0.5,0.5,1) and $\lambda=3$
it is of order unity, more precisely  1.4, which
is similar to the ratios found 
in the experiments of Refs.\ \cite{Mookpc,Keimer}.

\paragraph{Interlayer Coupling.} 
Earlier, we assumed that the order parameters from the
two layers in a bilayer are opposite in sign.
Let us see why this should be so.
Interactions which couple the density at a site
in one layer to the density at a site in
another layer will not couple the DDW
order parameters in the two layers
because they will average over directions in the $a-b$ plane,
giving zero. However, interlayer tunneling will couple
the DDW order parameters at second order. 

Consider the
interlayer tunneling Hamiltonian\cite{CSAS}  between layers (1) and (2):
$
H_c = -\sum_{{\bf k}\alpha}t_{\perp}({\bf k})
(c^{(1)\,\dagger}_{{\bf k}\alpha}c^{(2)}_{{\bf k}\alpha}+{\rm h. 
c.}), 
$
where $t_{\perp}({\bf k})=(t_{\perp}/4)(\cos k_x a-\cos k_y a)^2$.
 Let $|s\rangle$ be the ground
state for $t_{\perp}=0$ with equal DDW order parameters
in each layer, while $|a\rangle$ is
the ground state for $t_{\perp}=0$ with
DDW order parameters equal and opposite. The energy 
difference between these two
states, per site, is given to lowest non-vanishing order
in $t_\perp$ by
$
\frac{\Delta_{\rm sas}}{N}\approx -\frac{25}{32}
\frac{t_{\perp}^2}{\Delta_0}\Phi_{\bf Q}^2, 
$
where ${\Delta_0}=\sqrt{(\Delta^{\rm DDW})^2+(\Delta^{\rm DSC})^2}$,
and $\Delta^{\rm DSC}$ is the maximum of the superconducting
gap, i.e.  the antisymmetric state has lower energy. The above result also holds  for  body-centered tetragonal
materials, with the modified tunneling matrix element
\cite{vanderMarel}. Moreover, we see why the coupling between bilayers is so small:
the tunneling matrix elements between bilayers are much smaller
than those within a bilayer because
$c=11.82 \mbox{\AA} \gg d = 3.25 \mbox{\AA}$.

Because the coupling between the unit cells
is exceptionally small, we
expect to see $2D$ Bragg rods without $3D$ order for temperatures
which are not exceedingly low; recall that the transition
to the DDW state is an Ising transition unlike the corresponding
spin problem. Indeed, as remarked earlier,
Mook {\em et al.}\cite{Mook}
find such Bragg rods.

\paragraph{Inelastic scattering.} 
We note that because DDW order is
Ising-like, there are no
Goldstone modes, but  a gap in the frequency spectrum.
This is consistent with the experimental finding
of a spin gap above the elastic peak\cite{Mook}, extending up
to an energy of 20 meV.
For spin-ordering, a Bragg peak will necessarily lead
to a Goldstone mode (the Ising
anisotropy in these materials is negligibly small;
see, for example, Ref.~\cite{Shamoto}.), which is continuously
connected to it. It is not possible to have one without the other.

The magnetic resonance peak\cite{resonance} in inelastic scattering
at 34 meV at the same momentum, $\bf Q$,
also testifies against the spin-ordering scenario.
If the spins are ordered, this resonance should
be at zero energy -- i.e. it should not be distinct
from the Bragg peak with the same quantum numbers!
Furthermore, in the presence
of broken spin-rotational symmetry, excitations are
not organized into SU(2) multiplets (but, rather, U(1)
multiplets if rotations about one axis are preserved),
so the resonance could not possibly be a triplet mode as suggested 
in Ref.~\cite{Keimer}.

\paragraph{Global Phase Diagram.}

According to the interpretation proposed here,
DDW order has been observed in neutron
scattering in YBCO$_{6.6}$ \cite{Mook}.
On the other hand,
neutron scattering in YBCO$_{6.35}$ produces
a signal which is $10$ times more intense,
is due to moments which lie in the $a-b$ plane
and is believed to be due to
spin glass ordering\cite{Mook}.
This not only supports the hypothesis
that the effect observed in YBCO$_{6.6}$ is due to something
other than spin ordering -- namely DDW
ordering -- but also suggests that
DDW order disappears at low doping in favor of
spin glass order and, possibly, other competing
orders as well\cite{Kivelson}.

In neutron scattering
experiments on YBCO$_{6.5}$, Sidis {\em et al.} \cite{Keimer} find
similar magnetic Bragg peaks with
a ratio of $0.67$ between the
intensity at $(0.5, 0.5, 1)$ and that at
$(0.5, 0.5, 2)$. This is intermediate
between the observed ratio in
the putative DDW state
of YBCO$_{6.6}$ and the observed ratio in
the spin glass of YBCO$_{6.35}$. 
The measurements of Ref.~\cite{Keimer} 
seem to be indicative of the spin glass phase. {\em It is clear therefore that
the doping dependence may be crucial, changing the results as we move from 
YBCO$_{6.5}$ to YBCO$_{6.6}$.}

\begin{figure}[htb]
\centerline{\epsfxsize=2in\epsffile{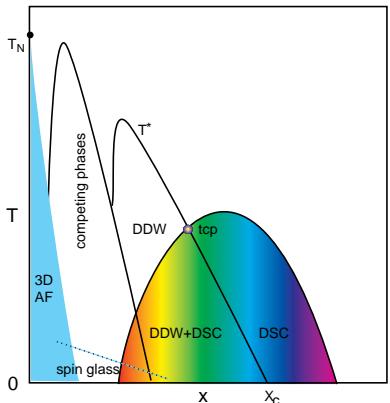}}
\caption{Phase diagram. 3D AF is the
three-dimensional antiferromagnetic phase, and $T_N$
is the corresponding N{\'e}el temperature. 
Competing phases refer to
a complex set of charge-ordered states;
$tcp$ is a tetracritical point; $x_c$ is a quantum
critical point, and $T^*$ signifies the pseudogap transition.} 
\label{fig:phase}
\end{figure}

With the above ideas as inspiration, we
propose the global phase diagram of
Fig.~\ref{fig:phase}.

\paragraph{Summary.}
To summarize, the ratio of spin-flip to non-spin-flip
scattering depends on details of the current distribution,
but the rapid decrease of the scattering intensity with
${\bf q}$ is a robust feature which follows from the
inherent spatial extent of the current loops
(which is much larger than a Cu or even an O
orbital).  We believe that DDW order 
is the only way of reconciling the constraints
following from the wavevector and polarization dependence
of the neutron scattering intensity. These phenomenological
considerations, although quite tight, clearly await a microscopic understanding.

{\em Note added:} Recent striking zero-field muon spin relaxation measurements
in YBCO$_{6.67}$ and YBCO$_{6.95}$ have uncovered the onset of static
magnetism consistent with our DDW picture, and the phase diagram proposed 
above\cite{muSR}, in particular, a phase transition within the 
superconducting dome.

S. C. is supported by NSF-DMR-9971138, and C. N. by 
NSF-DMR-9983544
and the A.P. Sloan Foundation.
The work of H. -Y. K. was conducted under
the auspices of the DOE, supported by funds
provided by the University of California for the conduct
of discretionary research by Los Alamos National Laboratory.
We thank E. Abrahams, B. Keimer, S. Kivelson, R. B. Laughlin,
and J. Tranquada for discussions. We
especially thank H. Mook and J. E. Sonier for interesting and important correspondence.

\end{document}